\documentstyle[12pt]{article}
\newlength{\dinwidth}
\newlength{\dinmargin}
\newlength{\extraspace}
\newlength{\extraspaces}
\newcommand{\be}{\begin{equation}
\addtolength{\abovedisplayskip}{\extraspaces}
\addtolength{\belowdisplayskip}{\extraspaces}
\addtolength{\abovedisplayshortskip}{\extraspace}
\addtolength{\belowdisplayshortskip}{\extraspace}}
\newcommand{\ee}{\end{equation}}
\newcommand{\bdm}{\begin{displaymath}
\addtolength{\abovedisplayskip}{\extraspaces}
\addtolength{\belowdisplayskip}{\extraspaces}
\addtolength{\abovedisplayshortskip}{\extraspace}
\addtolength{\belowdisplayshortskip}{\extraspace}}
\newcommand{\edm}{\end{displaymath}}
\renewcommand{\thefootnote}{\fnsymbol{footnote}}
\def\simlt{\mathrel{\lower2.5pt\vbox{\lineskip=0pt\baselineskip=0pt
           \hbox{$<$}\hbox{$\sim$}}}}
\def\simgt{\mathrel{\lower2.5pt\vbox{\lineskip=0pt\baselineskip=0pt
           \hbox{$>$}\hbox{$\sim$}}}}
\catcode`@=11
\newcount\@tempcntc
\def\@citex[#1]#2{\if@filesw\immediate\write\@auxout{\string\citation{#2}}\fi
  \@tempcnta\z@\@tempcntb\m@ne\def\@citea{}\@cite{\@for\@citeb:=#2\do
    {\@ifundefined
       {b@\@citeb}{\@citeo\@tempcntb\m@ne\@citea\def\@citea{,}{\bf ?}\@warning
       {Citation `\@citeb' on page \thepage \space undefined}}%
    {\setbox\z@\hbox{\global\@tempcntc0\csname b@\@citeb\endcsname\relax}%
     \ifnum\@tempcntc=\z@ \@citeo\@tempcntb\m@ne
       \@citea\def\@citea{,}\hbox{\csname b@\@citeb\endcsname}%
     \else
      \advance\@tempcntb\@ne
      \ifnum\@tempcntb=\@tempcntc
      \else\advance\@tempcntb\m@ne\@citeo
      \@tempcnta\@tempcntc\@tempcntb\@tempcntc\fi\fi}}\@citeo}{#1}}
\def\@citeo{\ifnum\@tempcnta>\@tempcntb\else\@citea\def\@citea{,}%
  \ifnum\@tempcnta=\@tempcntb\the\@tempcnta\else
   {\advance\@tempcnta\@ne\ifnum\@tempcnta=\@tempcntb \else \def\@citea{--}\fi
    \advance\@tempcnta\m@ne\the\@tempcnta\@citea\the\@tempcntb}\fi\fi}
\catcode`@=12

\newcommand{\la}{\lambda}

\newcommand{\bear}{\begin{eqnarray}}
\newcommand{\eear}{\end{eqnarray}}
\newcommand{\ba}{\begin{array}}
\newcommand{\ea}{\end{array}}

\newcommand{\trace}{\mbox{\rm Tr}}

\newcommand{\Dmu}{\mbox{$D^\mu$}}
\newcommand{\dmu}{\mbox{$D_\mu$}}
\newcommand{\pdmu}{\mbox{$\partial_\mu$}}

\newcommand{\ew}{\mbox{$SU(2)_{W}\times U(1)_{Y}$}}

\newcommand{\cB}{{\cal B}} \newcommand{\cO}{{\cal O}}

 \newcommand{\cX}{{\cal X}}
\newcommand{\cL}{{\cal L}}

\begin{document}
\begin{titlepage}
\begin{flushright}
BU-HEP 94-20\\
hep-ph/9409217\\
\today
\end{flushright}
\vspace{24mm}
\begin{center}
\Large{{\bf Comment on ``The Phenomenology of a Nonstandard Higgs Boson
in $W_L W_L$ Scattering''}}
\end{center}
\vspace{5mm}
\begin{center}
Dimitris Kominis\footnote{e-mail address: kominis@budoe.bu.edu}
\ \  and \ \
Vassilis Koulovassilopoulos\footnote{e-mail address:
vk@budoe.bu.edu}\\*[3.5mm]
{\normalsize\it Dept. of Physics, Boston University, 590 Commonwealth
Avenue,}\\
{\normalsize\it Boston, MA 02215}
\end{center}
\vspace{2cm}
\thispagestyle{empty}
\begin{abstract}
We show that in Composite Higgs models, the coupling of the Higgs
resonance to a pair of $W$ bosons is weaker than the corresponding
Standard Model coupling, provided the Higgs arises from electroweak
doublets only. This is partly due to the effects of the nonlinear
realization of the chiral symmetries at the compositeness scale.
\end{abstract}
\end{titlepage}
\newpage
\renewcommand{\thefootnote}{\arabic{footnote}}
\setcounter{footnote}{0}
\setcounter{page}{2}

In a recent paper \cite{vk}, Koulovassilopoulos and Chivukula presented a
Composite Higgs model based on the chiral symmetry breaking pattern
$SU(4)/Sp(4)$, where the coupling of the isoscalar ``Higgs'' resonance
to a pair of $W$ bosons (and consequently its partial decay width to
$WW$ and $ZZ$) was smaller than its value in the Standard Model.
In this Comment we show that, to lowest order in chiral perturbation
theory, this is true in all Composite Higgs models \cite{chm1,chm},
provided the isoscalar resonance arises from electroweak doublets only.
This fact is well known in the case of {\em linear} models of elementary
scalars. The new
element in the proof that follows is that the effects of the nonlinear
realization of the chiral symmetry
reduce the strength of the coupling of the
Higgs particle to $WW$ even further.

In Composite Higgs models, the Higgs arises as a pseudo-Goldstone boson
of the spontaneous breakdown of the chiral symmetries of ultrafermions.
We denote the chiral symmetry group by $G$. At some scale $f$, the
strong ultracolor dynamics causes the group $G$ to spontaneously break
down to a subgroup $H$. The Goldstone boson manifold $G/H$ can be
parametrized by the field
\be
\Sigma = \exp \left( \frac{2\,i\, \Pi^\alpha \cX^\alpha}{f}\right)
\ee
where $\Pi^\alpha$ are the Goldstone boson fields and $\cX^\alpha$ the
broken generators normalized so that $\trace\, (\cX^a \cX^b)=\frac{1}{2}
\delta^{ab}$. Under $g\in H$,  $\Sigma$ transforms as
\be
\Sigma \rightarrow g \, \Sigma \, g^{\dagger}  .
\ee

Since $\ew \subseteq H$, the interactions of the electroweak gauge bosons
with $\Pi^\alpha$ are described to lowest order in momentum by a chiral
lagrangian
\be
\cL_\Sigma = \frac{f^2}{4} \trace\, \left(\dmu\Sigma^\dagger\,\Dmu\Sigma
     \right) \label{eq:lagxi}
\ee
where the covariant derivative is 
\be
\dmu\Sigma = \pdmu\Sigma + i g [S^a,\Sigma] W^a_\mu
     +  i g^\prime [Y, \Sigma]\; \cB_{\mu}
\ee
where $S^a , Y$ belong to the algebra $H$ and generate $SU(2)_L$,
$U(1)_Y$ transformations respectively.

Among the $\Pi^\alpha$, there are four fields $\sigma, w_1, w_2, w_3$,
which
{\em by assumption} transform in the fundamental representation of
$SU(2)_L$, {\it i.e.} they form an electroweak doublet
\be
\Phi = \frac{1}{\sqrt{2}} \left ( \begin{array}{c} w_1+iw_2 \\
\sigma+iw_3 \end{array} \right )
\label{eksi}
\ee
In order to illustrate more clearly the
effects of the nonlinear realization,
we assume that only one (composite) doublet is
responsible for electroweak symmetry breaking.
In what follows, we shall set to zero all the other
Goldstone bosons since they do not affect our argument. Furthermore, we
can ignore also hypercharge ({\it i.e.} set $g^\prime =0$). The
generalization to $g^\prime \neq 0$ is straightforward.

The dynamics of the vacuum alignment is responsible for giving $\Sigma$
a vacuum expectation value $\langle\Sigma\rangle = \exp ( 2i G \langle
\sigma\rangle /f)$, where $G$ is the corresponding generator.
The term in the lagrangian (\ref{eq:lagxi}) which describes the
interactions of the Higgs with the gauge boson is
\be
\cL_{WW\sigma} = \frac{g^2}{8} M^2(\sigma)\: W_\mu^a W^{a \mu}
\ee
where
\be
\frac{1}{2} M^2(\sigma) = -f^2 \trace \left[S^3, e^{-\frac{2 i \sigma
  G}{f}}\right] \left[S^3, e^{\frac{2 i \sigma G}{f}}\right]
\label{eq:mw}
\ee
Note that $M^2(\sigma)$ is positive definite due to the
hermiticity of $S^3, G$. We have also set the exact Goldstone bosons $w_i$
to zero, by going over to the unitary gauge. By defining the shifted field
$\sigma = \langle\sigma\rangle + H$, the lagrangian above expanded in
terms of $H$  gives
\be
\cL_{WW\sigma} = \frac{1}{2} M^2_W W_\mu^a W^{a \mu} +
 \frac{g^2 v}{4} \xi H W_\mu^a W^{a \mu} + \cO (H^2)
\ee
where
\begin{eqnarray}
M^2_W  &= & \frac{1}{4} g^2 M^2(\langle\sigma\rangle) \\
v & = & M(\langle\sigma\rangle)
\end{eqnarray}
and
\be
\xi = M^\prime (\langle\sigma\rangle)
\ee
with the prime denoting differentiation. Thus, $\xi$ parametrizes the
strength of the Higgs coupling to a pair of $W$ bosons. The Standard
Model has $\xi=1$; hence we have to show that $M^\prime(\sigma) \leq 1$.

Consider first the limit $f \rightarrow \infty$. Then
\be
\frac{1}{2} M^2 (\sigma) = -4 \sigma^2 \trace\: [S^3,G] [S^3,G] + \cO\,
(1/f^2)
\label{eq:forio}
\ee
In order to evaluate this trace we have to make use of the assumption
that $\sigma$ belongs to the doublet (\ref{eksi}).
This is tantamount to the relation
\be
[S^3,G]= i \frac{1}{2} \cX^3
\ee
where $\cX^3$ is the broken generator that  corresponds to the $w_3$
Goldstone boson, correctly normalized $\trace\, (\cX^3)^2 =1/2$. It thus
follows that
\be
\trace\: [S^3,G] [S^3,G] = - \frac{1}{8} \;\;\; .  \label{eq:ft}
\ee
Consequently, from eq.~(\ref{eq:forio}) it follows that
$M (\sigma) = \sigma$
and thus $\xi=1$.  This is the limit where the composite Higgs models
reduce to the standard model. Notice however, that the
relation~(\ref{eq:ft}) holds even in the general case of finite $f$ which
we now consider.

To evaluate the expression (\ref{eq:mw}), it is
convenient to express $S^3$ as a sum of eigenvectors of~$G$:
\be
S^3 = \sum_{i} b_i E_i \label{eq:tbe}
\ee
where the $E_i$ are defined by
\be
[G, E_i]= \lambda_i E_i .  \label{eq:gee}
\ee
So $\lambda_i$ are the (real) eigenvalues of $G$ in the adjoint
representation.
If we normalize the $E_i$'s so that
\be
\trace\,E^\dagger_i E_j = \frac{1}{2} \delta_{ij}
\ee
then the normalization of $S^3$ implies that 
\be
\sum_i |b_i|^2 =1
\label{triavta}
\ee
while eq.~(\ref{eq:ft}) implies that
\be
\sum_i \lambda_i^2 \, |b_i|^2 = \frac{1}{4} \;\;\;\; .
\label{triavtaeva}
\ee

We are now in a position to evaluate eq.~(\ref{eq:mw}).
Let $U=\exp (2i\sigma G/f)$. Then
\be
\trace\: [S^3, U^\dagger][S^3, U] = 2 \,\trace\:
(S^3 U^\dagger S^3 U) -1 \;\; .
\label{eq:tutu}
\ee
By expanding $U$, and using the Baker-Campbell-Hausdorff formula and
eqs.~(\ref{eq:tbe}), (\ref{eq:gee}) we obtain
\bear
U^\dagger S^3 U &=& S^3 - \frac{2 i \sigma}{f} [G,S^3] + \frac{1}{2 !}
\left(-
 \frac{2 i \sigma}{f} \right)^2 \left[G,[G,S^3]\right] + \ldots \nonumber\\
&=& \sum_i b_i E_i \,e^{- 2 i \sigma\lambda_i/f}
\eear
Consequently,
\be
\trace\: (S^3 U^\dagger S^3 U) = \frac{1}{2} \sum_i |b_i|^2\, e^{-2i\sigma
\lambda_i/f}
\ee
Since ${\rm tr}\, S^3U^{\dagger}S^3U$ is real, it follows that
\be
{\rm tr}\, S^3U^{\dagger}S^3U = \frac{1}{2} \sum_i |b_i|^2 \cos \frac{2\sigma
\la_i}{f}
\label{triavtatessera}
\ee
and, by virtue of (\ref{triavta}),
\begin{eqnarray}
2\:{\rm tr}\, S^3U^{\dagger}S^3U -1 & = & \sum_i |b_i|^2 \left ( \cos \frac{2
\sigma \la_i}{f}-1 \right ) \nonumber \\
 & = & -2 \:\sum_i |b_i|^2 \sin ^2 \frac{\sigma \la_i}{f}
\label{triavtapevte}
\end{eqnarray}
Hence
\be
M^2(\sigma)=4\,f^2 \,\sum_i |b_i|^2 \sin ^2 \frac{\sigma \la_i}{f}
\ee
To show $M'(\sigma) \leq 1$, it suffices to show that
$(d/d\sigma)M^2(\sigma) \leq 2M(\sigma)$ (for $M>0$).
\begin{eqnarray}
\frac{d}{d\sigma} M^2(\sigma) & = & 8 f^2 \sum_i |b_i|^2 \,
\sin \frac{\sigma
\la_i}{f} \cos \frac{\sigma \la_i}{f} \cdot \frac{\la_i}{f} \nonumber \\
 & \leq & 8 f \sum_i |b_i|^2 |\la_i|  \left |\sin \frac{\sigma
\la_i}{f}\right |  \nonumber \\
 & \leq & 8 f \cdot  \sqrt{\,\sum_i \la_i^2 |b_i|^2} \:
\sqrt{\,\sum_i |b_i|^2
\sin ^2 \frac{\sigma \la_i}{f}} \nonumber \\
 & = & 2\, M(\sigma)
\end{eqnarray}
where we have used (\ref{triavtaeva}). The penultimate step is obvious
if we define ``vectors'' $\vec{A}, \vec{B}$ with components $A_i=|\la_i
b_i|, B_i=|b_i \sin (\sigma \la_i /f)|$ and employ the
fact that $\vec{A} \cdot \vec{B} \leq |\vec{A}| |\vec{B}|$.
This completes the proof that, to lowest order in chiral perturbation
theory, the parameter $\xi$ cannot exceed its Standard Model value of 1,
provided the Higgs belongs to an electroweak doublet representation. In
some Composite Higgs models \cite{chm1}, including the one presented in
ref.~\cite{vk}, the Higgs field appears as a linear combination of
fields that belong to two different doublets, one of which is usually
taken to be fundamental. In this case $\xi$ is reduced even further, in
the same way as in linear two-doublet models. For example, in the model
of ref.~\cite{vk}, $\xi$ can be written in terms of mixing angles
$\alpha$ and $\beta$ as
\be
\xi = \sin\alpha \sin\beta + M^\prime (\sigma) \cos\alpha \cos\beta
\ee
which obviously is smaller than one since $M^\prime (\sigma)\leq 1$.

We thus conclude that, if the Higgs resonance arises from electroweak
doublets only, then its
coupling to a pair of $W$ bosons is smaller than the corresponding
Standard Model coupling. In order to obtain $\xi > 1$, the Higgs boson
must be part of a
larger representation of $SU(2)_L$. This possibility has
been emphasized recently by Chivukula, Dugan and Golden \cite{cdg}.

We thank R.~S.~Chivukula and R.~Rohm for useful discussions,
and M.~Golden and M.~Dugan for reading the manuscript. This work was
supported in part under NSF contract PHY-9057173 and DOE contract
DE-FG02-91ER40676.

\end{document}